\documentclass[preprint,eqsecnum,aps,nofootinbib]{revtex4}
\usepackage{amsfonts,amsmath,amssymb,amsthm}
\usepackage{latexsym}
\usepackage{bbm,bm}
\usepackage{graphicx}

\newcommand{\ket}[1]{\lvert #1 \rangle}
\newcommand{\bra}[1]{\langle #1 \lvert}
\newcommand{\beq}{\begin{equation}}
\newcommand{\eeq}{\end{equation}}
\newcommand{\beqs}{\begin{eqnarray}}
\newcommand{\eeqs}{\end{eqnarray}}

\begin{document}

\title{Entanglement Classification of extended Greenberger-Horne-Zeilinger-Symmetric States}

\author{Eylee Jung$^{1}$ and DaeKil Park$^{1,2}$}

\affiliation{
             $^1$Department of Electronic Engineering, Kyungnam University, Changwon
                 631-701, Korea       \\
             $^2$Department of Physics, Kyungnam University, Changwon
                  631-701, Korea                                            
             }

\begin{abstract}
In this paper we analyze entanglement classification of extended Greenberger-Horne-Zeilinger-symmetric states $\rho^{ES}$, which is parametrized
by four real parameters $x$, $y_1$, $y_2$ and $y_3$. The condition for separable states of $\rho^{ES}$ is analytically derived. 
The higher classes such as bi-separable, W, and Greenberger-Horne-Zeilinger classes are roughly classified by making use of 
the class-specific optimal witnesses or map from the extended Greenberger-Horne-Zeilinger symmetry to the Greenberger-Horne-Zeilinger symmetry. 
From this analysis we guess that the entanglement classes of $\rho^{ES}$ are not dependent on $y_j \hspace{.2cm} (j=1,2,3)$ individually, 
but dependent on $y_1 + y_2 + y_3$ collectively. The difficulty arising in extension of analysis with 
Greenberger-Horne-Zeilinger symmetry to the higher-qubit system is discussed.
\end{abstract}

\maketitle

\section{Introduction}
Entanglement\cite{horodecki09} is an important physical resource in the context of quantum information theories\cite{text}. 
As shown for last two decades it plays a crucial role in quantum teleportation\cite{teleportation}, superdense coding\cite{superdense},
quantum cloning\cite{clon}, quantum cryptography\cite{cryptography}. It is also quantum entanglement, which makes the quantum
computer outperform the classical one\cite{computer}. Therefore, it is greatly important task to understand what kind and 
how much entanglement a given quantum state has.

For multipartite quantum states there are several types of entanglement. Each type is in general categorized by 
stochastic local operations and classical communication (SLOCC)\cite{bennet00}. Thus, these types of entanglement is 
often called SLOCC-equivalence classes. For example, for three-qubit pure states\cite{dur00} 
there are six SLOCC-equivalence classes such as separable, three bi-separable ($A-BC$, $B-AC$, $C-AB$), W and 
Greenberger-Horne-Zeilinger (GHZ) classes. Among them genuine tripartite entanglement arises in W and GHZ classes. 
The representative states of these classes are 
\begin{eqnarray}
\label{representative}
& &\ket{\mbox{GHZ}} = \frac{1}{\sqrt{2}} \left[ \ket{000} + \ket{111} \right]               \\    \nonumber
& &\ket{\mbox{W}} = \frac{1}{\sqrt{3}} \left[ \ket{001} + \ket{010} + \ket{100} \right].
\end{eqnarray}
One of the most remarkable fact in this classification is that the set of W states forms measure zero in the whole three-qubit
pure states.

This classification can be extended to three-qubit mixed states\cite{threeM}. Following 
Ref.\cite{threeM} the whole three-qubit mixed states are classified as separable (S), bi-separable (B),
W and GHZ classes. These classes satisfy a linear hierarchy S $\subset$ B $\subset$ W $\subset$ GHZ. 
One remarkable fact, which was proved in this reference, is that the
W-class\footnote{As Ref. \cite{elts12-2} we will use the names of the SLOCC classes in an exclusive sense throughout this paper.} 
is not of measure zero among all mixed-states.

Although SLOCC classes for three-qubit system are well-known, we still do not know how the entanglement is 
classified in multi-qubit system except four-qubit pure states, where there are nine SLOCC classes\cite{fourP}.
Furthermore, still it is very difficult problem to find a SLOCC class of a given three-qubit mixed state\footnote{For three-qubit
pure states it is possible to find the SLOCC classes by computing the concurrence\cite{concurrence1} of the reduced 
states and three-tangle\cite{ckw} for the given states.} except few rare cases. Thus, it is important task
to develop a method, which enables us to find a SLOCC class of an arbitrary three-qubit states.

Recently, a significant progress is made in this issue. In Ref.\cite{elts12-1} a complete SLOCC classification for the set of the 
GHZ-symmetric states was reported (see Fig. 1). According to this complete classification the ratio of number of S, B, W, and GHZ states
in the whole set of the GHZ-symmetric states is $1:1:1.076:0.924$. Thus, W class is not of measure zero in this restricted set of 
the three-qubit states. Using this classification the three-tangle $\tau$ for the arbitrary GHZ-symmetric states is explored 
in Ref.\cite{siewert12-1}. Moreover, this complete classification is used to construct the class-specific optimal witnesses for the 
three-qubit entanglement\cite{elts12-2}.

The purpose of this paper is to explore a possible extension of Ref.\cite{elts12-1} to treat more three-qubit mixed states. For this 
purpose we enlarge the symmetry group to, so-called, the extended GHZ symmetry group. The whole set of quantum states
invariant under the extended GHZ symmetry group is parametrized by four real parameters $(x, y_1, y_2, y_3)$. The complete classification
for S states is analytically derived. However, the classification for B, W, and GHZ states is incomplete. Rough classification
for B, W, and GHZ states is explored by making use of the class-specific optimal witnesses\cite{elts12-2} or 
GHZ symmetry\cite{elts12-1}, respectively.

The paper is organized as follows. In next section we review Ref. \cite{elts12-1} briefly. In section III we discuss on the 
extended GHZ symmetry. It is found that the set of extended GHZ-symmetric states is parametrized by four real parameters. In section IV
we derive a condition for the separable region in the four-dimensional parameter space by applying a Lagrange multiplier method. 
In section V we perform a entanglement classification of the extended GHZ-symmetric states roughly by making use of 
the class-specific optimal witnesses or map from extended GHZ symmetry to GHZ symmetry. In section VI a conclusion is given. In particular,
we discuss on the difficulty arising when we extend the analysis with the GHZ symmetry to the higher-qubit systems in this section.

\section{classification of GHZ-symmetric states}
In this section we review Ref. \cite{elts12-1} briefly. The GHZ-symmetric states are the three-qubit states which are 
invariant under the following transformations: (i) qubit permutations, (ii) simultaneous three-qubit flips (i.e., application of 
$\sigma_x \otimes \sigma_x \otimes \sigma_x$), (iii) qubit rotations about the $z$-axis of the form
\begin{equation}
\label{z-rotation}
U (\phi_1, \phi_2) = e^{i \phi_1 \sigma_z} \otimes e^{i \phi_2 \sigma_z} \otimes e^{-i (\phi_1 + \phi_2) \sigma_z}.
\end{equation}

It is straightforward to show that the general form of the GHZ-symmetric states $\rho^S$ is parametrized by two real 
parameters $x$ and $y$ as 
\begin{eqnarray}
\label{ghz-general-form}
& &\rho^S (x, y) = \left( x + \frac{\sqrt{3}}{2} y + \frac{1}{8} \right) \ket{\mbox{GHZ}_+} \bra{\mbox{GHZ}_+} + 
            \left( -x + \frac{\sqrt{3}}{2} y + \frac{1}{8} \right) \ket{\mbox{GHZ}_-} \bra{\mbox{GHZ}_-}                  \\   \nonumber
& & + \left( \frac{1}{8} - \frac{y}{2 \sqrt{3}} \right)
\bigg[ \ket{001}\bra{001} + \ket{010}\bra{010} + \ket{011}\bra{011} + \ket{100}\bra{100} + \ket{101}\bra{101} + \ket{110}\bra{110} \bigg]
\end{eqnarray}
where
\begin{equation}
\label{ghz-states}
\ket{\mbox{GHZ}_{\pm}} = \frac{1}{\sqrt{2}} \left( \ket{000} \pm \ket{111} \right).
\end{equation}
The real parameters $x$ and $y$ are introduced such that the Euclidean metric in the $(x, y)$ plane coincides with the 
Hilbert-Schmidt metric $d (A, B)^2 \equiv \frac{1}{2} \mbox{tr} (A - B)^{\dagger} (A - B)$, i.e.,
\begin{equation}
\label{metric1}
d^2 \left( \rho^S (x_1, y_1), \rho^S (x_2, y_2) \right) = (x_2 - x_1)^2 + (y_2 - y_1)^2.
\end{equation}
Since $\rho^S (x, y)$ is a quantum state, the parameters $x$ and $y$ are restricted as 
\begin{equation}
\label{restrict1}
-\frac{1}{4 \sqrt{3}} \leq y \leq \frac{\sqrt{3}}{4}, \hspace{1.0cm} y \geq \pm \frac{2}{\sqrt{3}} x - \frac{1}{4 \sqrt{3}}.
\end{equation}
This restriction can be easily derived by computing the eigenvalues of $\rho^S$. Thus, the set of the GHZ-symmetric states are 
represented as a triangle in $(x, y)$ plane as Fig. 1 shows. Each point inside the triangle corresponds to each 
GHZ-symmetric state.

\begin{figure}[ht!]
\begin{center}
\includegraphics[height=8cm]{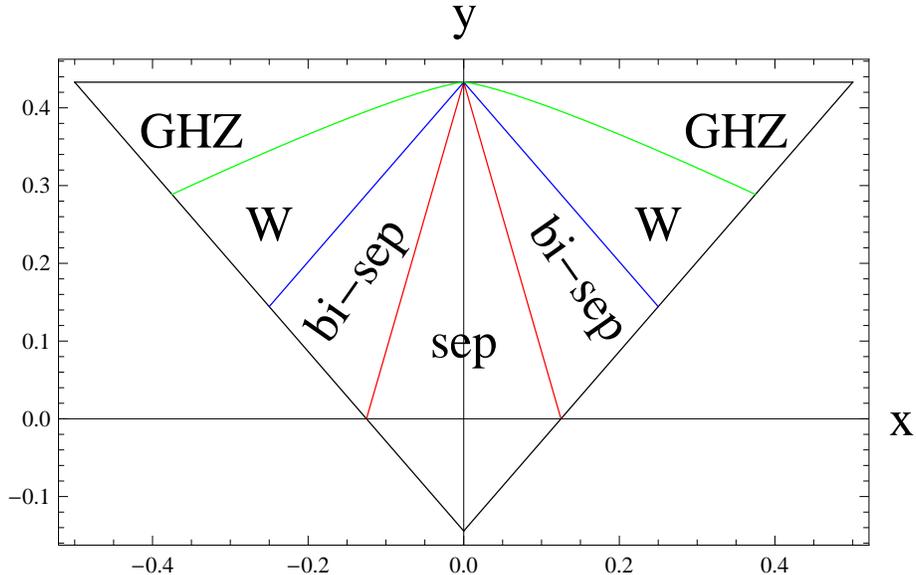}
\caption[fig1]{Complete classification of GHZ-symmetric states.}
\end{center}
\end{figure}

In order to classify the GHZ-symmetric states it is worthwhile noting that there exists a map from an arbitrary three-qubit 
pure state $\ket{\psi}$ to the GHZ-symmetric state $\rho^S (\psi)$, which is defined as 
\begin{equation}
\label{map1}
\rho^S (\psi) = \int d U U \ket{\psi} \bra{\psi} U^{\dagger},
\end{equation}
where the integral is understood to cover the entire GHZ symmetry group. If, for example, $\ket{\psi} = \sum_{i,j,k = 0}^1 \psi_{ijk} \ket{ijk}$,
the corresponding $\rho^S (\psi)$ is given by Eq. (\ref{ghz-general-form}) with 
\begin{eqnarray}
\label{symm-1}
x = \frac{1}{2} \left(\psi_{000}^* \psi_{111} + \psi_{000} \psi_{111}^* \right)                \hspace{1.0cm}
y = \frac{1}{\sqrt{3}} \left( |\psi_{000}|^2 + |\psi_{111}|^2 - \frac{1}{4} \right).
\end{eqnarray}
Another fact we will use for classification is that applying $GL(2,\mathbb{C})$ transformations to any qubit does not change 
the entanglement class of a multiqubit state. This fact leads from the invariance of the entanglement class under the stochastic local 
operations and classical communication (SLOCC)\cite{dur00,bennet00}.

In order to find the boundary of each entanglement class, therefore, we fix the $y$ coordinate and derive the maximum of $|x|$ by using
Eq. (\ref{symm-1}) and applying the Lagrange multiplier method. Since mirror symmetry implies $x_{min} = -x_{max}$, it is possible to 
restrict ourselves to $x \geq 0$. This procedure yields some region in the $(x,y)$ plane. If this region is a convex, the set of states 
corresponding to this region exhibits a same entanglement property. If it is not convex, the proper boundary of the class is obtained 
by the convex hull of this region.

The entanglement classification derived in this way is summarized in Fig. 1. Recently, this classification was used to compute the 
three-tangle\cite{ckw} of the entire GHZ-symmetric states analytically\cite{siewert12-1}. More recently, this is used to derive the 
class-specific optimal witnesses for three-qubit entanglement\cite{elts12-2}. 

\section{Extended GHZ symmetry}
In this section we will relax the condition of the GHZ symmetry to treat more large set of the three-qubit states. The symmetry we 
consider is identical with the GHZ symmetry without first condition, i.e., qubit permutations. We will call this the extended 
GHZ symmetry.

It is not difficult to show that the general form of the extended GHZ-symmetric states is parametrized by four real parameters 
$x$, $y_1$, $y_2$, and $y_3$ as 
\begin{eqnarray}
\label{eghz-general-form}
& &\rho^{ES} (x, y_1, y_2, y_3)                                                                                         \\   \nonumber                                                                                 
& & = \left(\frac{1}{8} + \frac{y_1 + y_2 + y_3}{2} + x \right) \ket{GHZ_+} \bra{GHZ_+} + 
      \left(\frac{1}{8} + \frac{y_1 + y_2 + y_3}{2} - x \right) \ket{GHZ_-} \bra{GHZ_-}                                  \\  \nonumber
& & +  \left(\frac{1}{8} - \frac{y_1 + y_2 - y_3}{2} \right) \bigg[ \ket{001} \bra{001} + \ket{110} \bra{110} \bigg]
+    \left(\frac{1}{8} - \frac{y_1 - y_2 + y_3}{2} \right) \bigg[ \ket{010} \bra{010} + \ket{101} \bra{101} \bigg]       \\   \nonumber
& & +    \left(\frac{1}{8} - \frac{-y_1 + y_2 + y_3}{2} \right) \bigg[ \ket{011} \bra{011} + \ket{100} \bra{100} \bigg].
\end{eqnarray}
The parameters are chosen so that the four-dimensional Euclidean metric coincides with the Hilbert-Schmidt metric, i.e.,
\begin{equation}
\label{metric2}
d^2 \bigg( \rho^{ES} (\bar{x}, \bar{y}_1, \bar{y}_2, \bar{y}_3), \rho^{ES} (x, y_1, y_2, y_3) \bigg)
= (\bar{x} - x)^2 + (\bar{y}_1 - y_1)^2 + (\bar{y}_2 - y_2)^2 + (\bar{y}_3 - y_3)^2.
\end{equation}

\begin{figure}[ht!]
\begin{center}
\includegraphics[height=6cm]{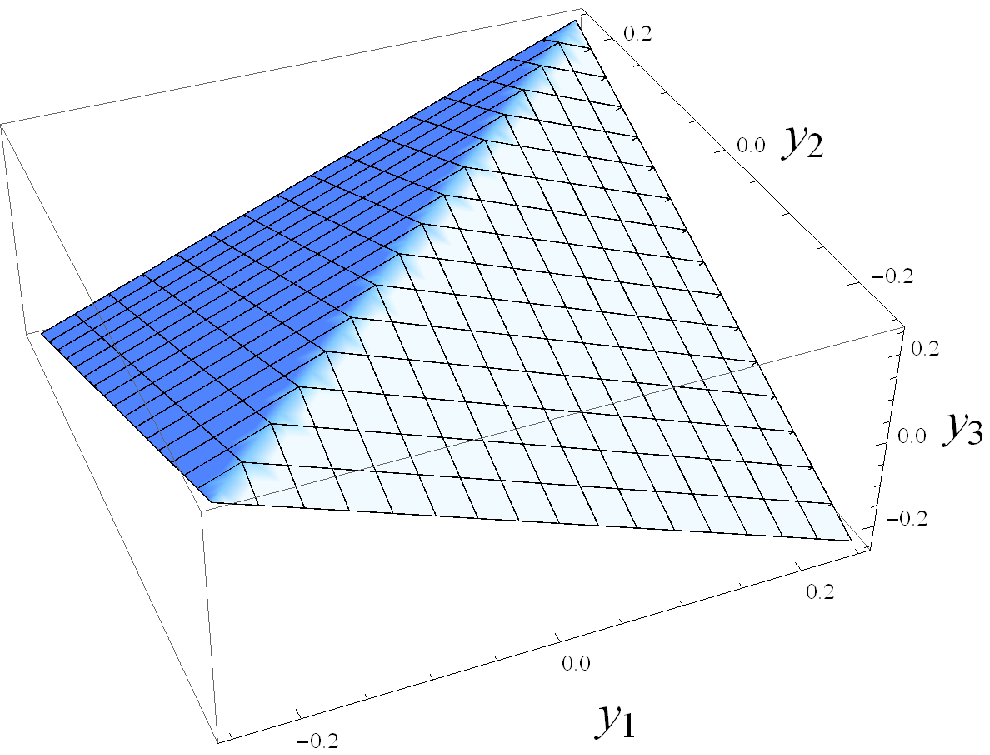}
\includegraphics[height=6cm]{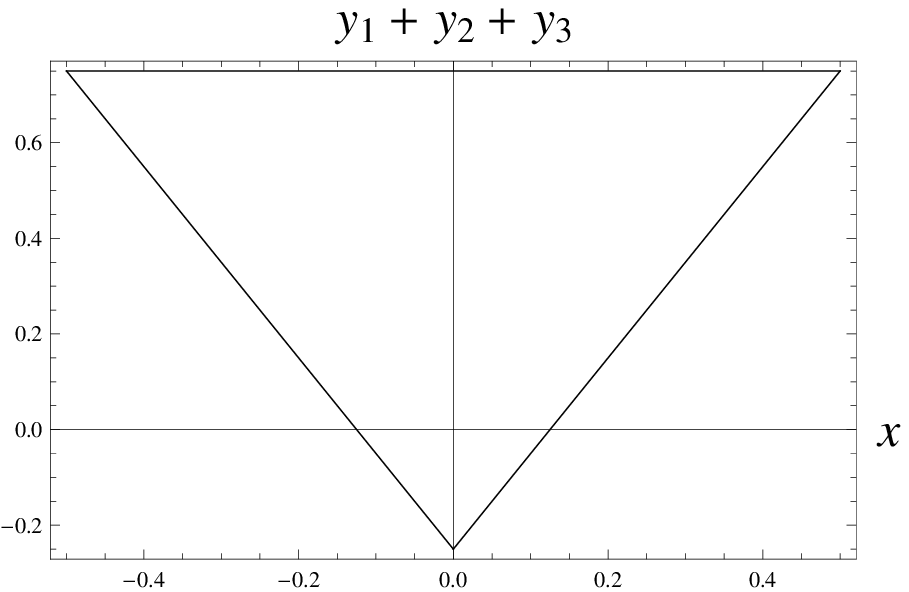}
\caption[fig2]{Pictorial representation of Eq. (\ref{restrict2}). Unlike the GHZ-symmetric case each point inside the triangle in (b)
corresponds to infinite number of quantum states with same $y_1+y_2+y_3$.}
\end{center}
\end{figure}

Since $\rho^{ES}$ should be a physical state, the parameters are restricted as 
\begin{equation}
\label{restrict2}
|y_1 + y_2| - \frac{1}{4} \leq y_3 \leq \frac{1}{4} - |y_1 - y_2|, \hspace{1.0cm}
0 \leq \frac{1}{8} + \frac{y_1 + y_2 + y_3}{2} \pm x \leq 1.
\end{equation} 
The restriction (\ref{restrict2}) can be depicted pictorially. As Fig. 2(a) shows, the physically available value of $y_i \hspace{.1cm} (i=1,2,3)$
is confined inside polyhedron in the three-dimensional $(y_1, y_2, y_3)$ space. As this figure exhibits, $y_i$'s are restricted by 
$-\frac{1}{4} \leq y_1, y_2, y_3 \leq \frac{1}{4}$. However, as Fig. 2(b) shows, $y_1 + y_2 + y_3$ is restricted by 
$-\frac{1}{4} \leq y_1 + y_2 + y_3 \leq \frac{3}{4}$ depending on $x$. Unlike the GHZ-symmetric case each point inside the triangle in Fig. 2(b)
corresponds to infinite number of quantum states with same $y_1+y_2+y_3$. 

Similarly to GHZ symmetry there exists a mapping from a set of the three-qubit pure states to the set of the extended GHZ-symmetric states.
Let $\ket{\psi}$ be an arbitrary three-qubit pure state. Then, the corresponding extended GHZ-symmetric state is given by Eq. (\ref{map1}).
The only difference is a change of the symmetry group from the GHZ symmetry to the extended GHZ symmetry. If, for example, 
$\ket{\psi} = \sum_{i,j,k=0}^{1} \psi_{ijk} \ket{ijk}$, the corresponding $\rho^{ES} (\psi)$ is given by Eq.(\ref{eghz-general-form}) with
\begin{eqnarray}
\label{symm-2}
& &x = \frac{1}{2} \left( \psi_{000} \psi_{111}^* + \psi_{000}^* \psi_{111} \right)                          \\    \nonumber
& &y_1 = \frac{1}{2} \left( |\psi_{000}|^2 + |\psi_{111}|^2 + |\psi_{011}|^2 + |\psi_{100}|^2 \right) - \frac{1}{4}     \\    \nonumber
& &y_2 = \frac{1}{2} \left( |\psi_{000}|^2 + |\psi_{111}|^2 + |\psi_{101}|^2 + |\psi_{010}|^2 \right) - \frac{1}{4}     \\    \nonumber
& &y_3 = \frac{1}{2} \left( |\psi_{000}|^2 + |\psi_{111}|^2 + |\psi_{110}|^2 + |\psi_{001}|^2 \right) - \frac{1}{4}.
\end{eqnarray}

It is worthwhile noting a relation 
\begin{equation}
\label{lu-3-1}
(u \otimes u \otimes u ) \rho^{ES} (x, y_1, y_2, y_3) (u \otimes u \otimes u )^{\dagger} = \rho^{ES} (-x, y_1, y_2, y_3)
\end{equation}
where $u = \left( \begin{array}{cc} 0  &  1  \\ -1  &  0  \end{array}  \right)$. This implies that the sign of $x$ does not change the 
entanglement class of $\rho^{ES} (x, y_1, y_2, y_3)$. Therefore, it is convenient to restrict ourselves to $x \geq 0$ in the following. 

\section{Separable states}
In this section we will find a region in the four-dimensional $(x, y_1, y_2, y_3)$ space, where the separable states reside. The 
calculation procedure is similar to Ref.\cite{elts12-1}. First, we define a general form of the fully separable three-qubit
pure state by making use of the local-unitary transformation, i.e., $\ket{\psi^{sep}} = \left( U_1 \otimes U_2 \otimes U_3 \right) \ket{000}$,
where
\begin{equation}
\label{sep-1}
U_j  =\left(          \begin{array}{cc}
                       A_j  &  -B_j^*        \\
                       B_j  &  A_j^*
                      \end{array}                 \right)
\hspace{1.0cm}  |B_j|^2 = 1 - |A_j|^2.
\end{equation}
Second, we map $\ket{\psi^{sep}}$ to the extend GHZ-symmetric state $\rho^{ES}(\psi^{sep})$ by using a map discussed in the previous section. 
Finally, we maximize $x$ when $y_1$, $y_2$, and $y_3$ are fixed.

Combining Eq. (\ref{symm-2}) and Eq. (\ref{sep-1}), $\rho^{ES}(\psi^{sep})$ is given by Eq. (\ref{eghz-general-form}) with
\begin{eqnarray}
\label{sep-2}
& &x = |A_1| |A_2| |A_3| \sqrt{1 - |A_1|^2} \sqrt{1 - |A_2|^2} \sqrt{1 - |A_3|^2}                        \\   \nonumber
& &\mu_1 = |A_2|^2 |A_3|^2 + (1 - |A_2|^2) (1 - |A_3|^2)                                                 \\   \nonumber
& &\mu_2 = |A_1|^2 |A_3|^2 + (1 - |A_1|^2) (1 - |A_3|^2)                                                 \\   \nonumber
& &\mu_3 = |A_1|^2 |A_2|^2 + (1 - |A_1|^2) (1 - |A_2|^2),
\end{eqnarray}
where $\mu_i = 2 y_i + \frac{1}{2}$.
In order to apply the Lagrange multiplier method we define
\begin{equation}
\label{sep-3}
x^{\Lambda} = x + \sum_{i = 1}^3 \Lambda_i \Theta_i,
\end{equation}
where $\Lambda_i$'s are the Lagrange multiplier constants and $\Theta_i$'s are the constraints given by 
\begin{eqnarray}
\label{sep-4}
& &\Theta_1 = |A_2|^2 |A_3|^2 + (1 - |A_2|^2) (1 - |A_3|^2) - \mu_1                                    \\   \nonumber
& &\Theta_2 = |A_1|^2 |A_3|^2 + (1 - |A_1|^2) (1 - |A_3|^2) - \mu_2                                    \\   \nonumber
& &\Theta_3 = |A_1|^2 |A_2|^2 + (1 - |A_1|^2) (1 - |A_2|^2) - \mu_3.
\end{eqnarray}

Before proceeding further, it is worthwhile to compare Eq. (\ref{sep-3}) with the corresponding equation derived for GHZ-symmetric
case at this stage.
For GHZ-symmetric case\cite{elts12-1} $x^{\Lambda}$ for the separable states becomes 
\begin{eqnarray}
\label{ghz-case-1}
& &x^{\Lambda} = |A_1| |A_2| |A_3| \sqrt{1 - |A_1|^2} \sqrt{1 - |A_2|^2} \sqrt{1 - |A_3|^2} + \Lambda \Theta     \\   \nonumber
& &\Theta = |A_1|^2 |A_2|^2 |A_3|^2 + (1 - |A_1|^2) (1 - |A_2|^2) (1 - |A_3|^2) - \left(\sqrt{3} y + \frac{1}{4} \right).
\end{eqnarray}
Thus $x^{\Lambda}$ in Eq. (\ref{ghz-case-1}) has a $A_i \leftrightarrow A_j$ symmetry. Thus, the maximum of $x$ occurs when
$|A_1| = |A_2| = |A_3|$, which drastically simplifies the calculation. However, as Eq. (\ref{sep-3}) and Eq. (\ref{sep-4}) show, 
$x^{\Lambda}$ in Eq. (\ref{sep-3}) does not have this symmetry. This is due to the fact that the extended GHZ symmetry is less 
symmetric than the GHZ symmetry.

The Lagrange multiplier method generates three equations $\frac{\partial x^{\Lambda}}{\partial |A_i|} = 0 \hspace{,2cm} (i=1,2,3)$. Since
we have three Lagrange multiplier constants, these equations can be used to express $\Lambda_i$ in terms of $|A_i|$. Thus, we should 
determine $|A_i|$ from only three constraints $\Theta_i = 0$, which yields
\begin{equation}
\label{sep-5}
|A_1|^2 = \frac{y_1 \pm 2 \sqrt{y_1 y_2 y_3}}{2 y_1}       \hspace{1cm}
|A_2|^2 = \frac{y_2 \pm 2 \sqrt{y_1 y_2 y_3}}{2 y_2}       \hspace{1cm}
|A_3|^2 = \frac{y_3 \pm 2 \sqrt{y_1 y_2 y_3}}{2 y_3}.       
\end{equation}
Therefore, $|A_1| = |A_2| = |A_3|$ doe not hold unless $y_1 = y_2 = y_3$. This is due to the less-symmetric nature of the extended 
GHZ symmetry compared to the GHZ symmetry.
From Eq. (\ref{sep-5}) $x_{max}$ is given by 
\begin{equation}
\label{sep-6}
x_{max} = \frac{1}{8}
\sqrt{\frac{(y_1 - 4 y_2 y_3) (y_2 - 4 y_1 y_3) (y_3 - 4 y_1 y_2)}{y_1 y_2 y_3}}.
\end{equation}
Eq. (\ref{sep-6}) gives a certain boundary in the four-dimensional $(x, y_1, y_2, y_3)$ space, inside of which the extended GHZ-symmetric 
separable states reside. It is worthwhile noting two points at the present stage. First, if the term inside the square root in 
r.h.s. of Eq. (\ref{sep-6}) is negative at some point (or region) inside the polyhedron of Fig. 2(a), this means that this point is excluded 
from the boundary. This is similar to $-\frac{1}{4 \sqrt{3}} \leq y < 0$ region in the GHZ-symmetric case as Fig. 1 exhibits. 
Second, if the region generated by Eq. (\ref{sep-6}) is not convex, we should extend it to its convex hull because the set of each entanglement 
class should be convex set. 

\begin{figure}[ht!]
\begin{center}
\includegraphics[height=5cm]{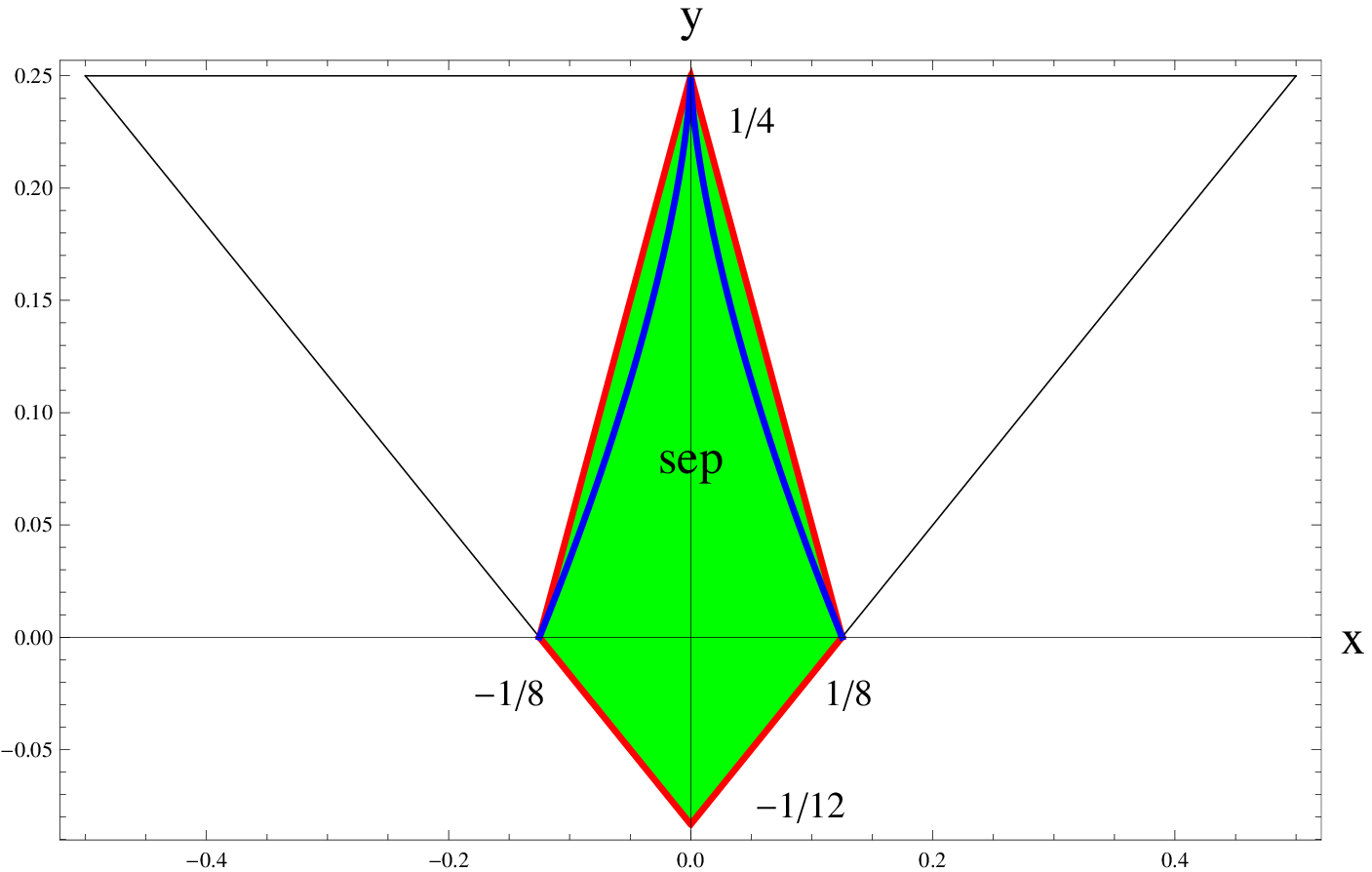}
\includegraphics[height=5cm]{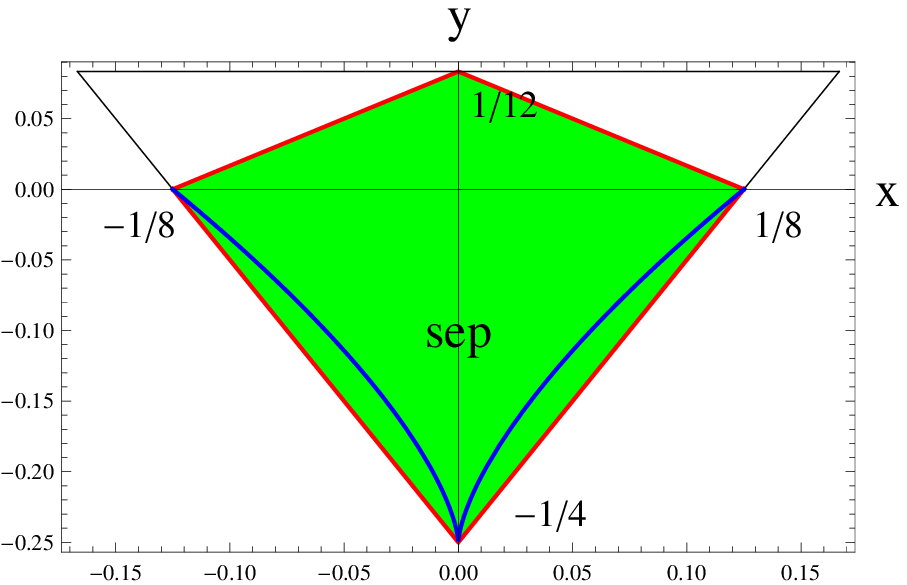}
\includegraphics[height=5cm]{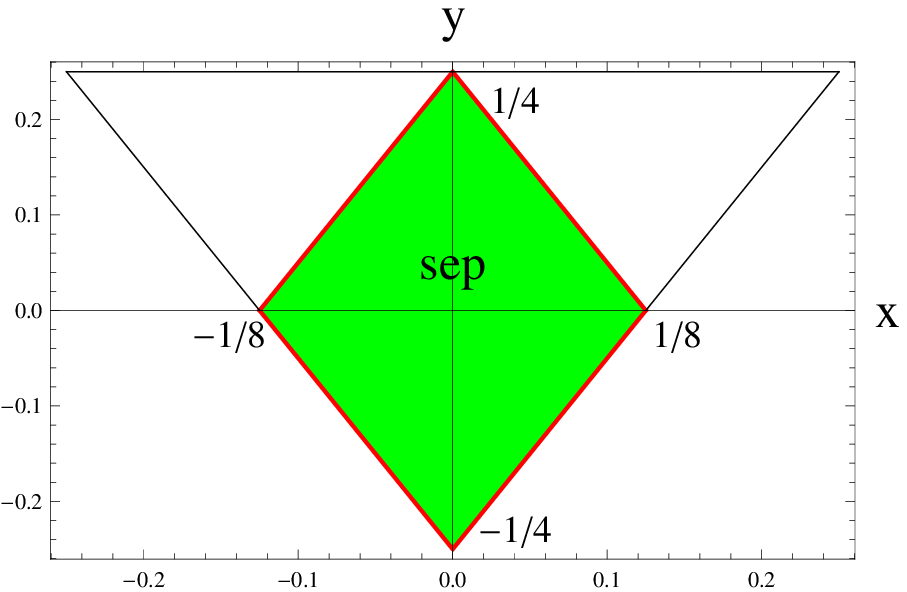}
\caption[fig3]{Separable region for (a) $y_1 = y_2 = y_3 \equiv y$, (b) $y_1 = y_2 = - y_3 \equiv y$, (c) $y_1 = y_2 = 0$ and $y_3 \equiv y$.
For first two cases Eq. (\ref{sep-6}) generates concave regions (see blue lines). Thus the convex hull (red line) for each case is chosen as a 
separable region. For last case, however, Eq. (\ref{sep-6}) generates a convex separable region (see red line). 
If $y_1 = -y_2 = -y_3 \equiv y$, the separable region in the $(x, y)$ plane becomes the same with (b) upside down.}
\end{center}
\end{figure}

Now, we consider several special cases. If $y_1 = y_2 = y_3 \equiv y$, Eq. (\ref{restrict2}) gives $-\frac{1}{12} \leq y \leq \frac{1}{4}$ and 
$x_{max}$ becomes
\begin{equation}
\label{sep-7}
x_{max} = \frac{1}{8} (1 - 4 y)^{3/2}.
\end{equation}
Since this is not convex (see blue line of Fig. 3(a)), we have to choose a convex hull, which is 
\begin{equation}
\label{sep-8}
x_{max} = \frac{1}{8} - \frac{y}{2}.
\end{equation}
This region is depicted in Fig. 3(a) as a green color.

As a second example, let us consider a case of $y_1 = y_2 = - y_3 \equiv y$. In this case Eq. (\ref{restrict2}) gives 
$-\frac{1}{4} \leq y \leq \frac{1}{12}$. Then, from Eq. (\ref{sep-6}) $x_{max}$ reduces to 
\begin{equation}
\label{sep-9}
x_{max} = \frac{1}{8} (1 + 4 y)^{3/2}.
\end{equation}
Since this is not convex (see blue line of Fig. 3(b)) and it is evident that the states with $x=0$ are separable, 
we should choose a separable region as a green color region in Fig. 3(b). 
If $y_1 = -y_2 = -y_3 \equiv y$, the separable region in the $(x, y)$ plane 
becomes the same with Fig. 3(b) upside down.

As a third example, let us consider a case of $(y_1, y_2, y_3) = (0, 0, y)$. In this case Eq. (\ref{restrict2}) gives 
$-\frac{1}{4} \leq y \leq \frac{1}{4}$ and Eq. (\ref{sep-6}) yields
\begin{equation}
\label{sep-11}
x_{max} = \frac{1}{8} (1 - 4 y).
\end{equation} 
The corresponding separable region is plotted in Fig. 3(c) as a green color. Since it is convex, we do not need to choose a convex hull in this case.

Finally, let us consider the positive partial transpose (PPT) states in the extended GHZ-symmetric states $\rho^{ES} (x, y_1, y_2, y_3)$.
Taking a partial transposition over the first qubit and computing the eigenvalues of the resulting matrix, one can derive the PPT condition 
for positive $x$ as $x \leq \alpha_4$, where
\begin{equation}
\label{sep-12}
\alpha_2 = \frac{1}{8} - \frac{y_1 + y_2 - y_3}{2},  \hspace{1.0cm}
\alpha_3 = \frac{1}{8} - \frac{y_1 - y_2 + y_3}{2},  \hspace{1.0cm}
\alpha_4 = \frac{1}{8} - \frac{-y_1 + y_2 + y_3}{2}.
\end{equation}
Performing similar calculation for second and third qubits, it is straightforward to derive a PPT condition of $\rho^{ES} (x, y_1, y_2, y_3)$
as
\begin{equation}
\label{sep-13}
x \leq x_{max} = \min (\alpha_2, \alpha_3, \alpha_4).
\end{equation}
One can show that the separable regions in Fig. 3 exactly coincide with the region, where the PPT condition (\ref{sep-13}) holds.

\section{Rough Classification using the class-specific optimal witness operators}

\begin{figure}[ht!]
\begin{center}
\includegraphics[height=10.0cm]{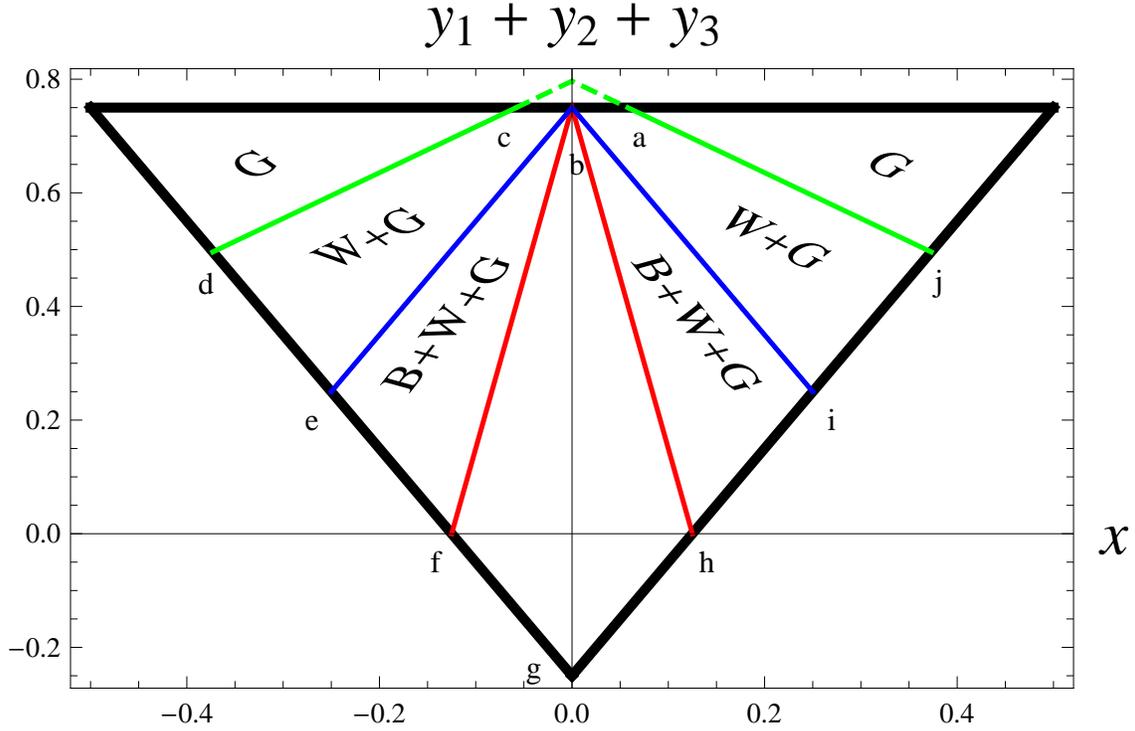}
\caption[fig4]{Rough SLOCC classification of the extended GHZ-symmetric states $\rho^{ES}$ given in Eq. (\ref{eghz-general-form}). 
G, W, and B stand for GHZ class, W class, and bi-separable class. The symbol `+' means coexistence. For example, 
W+G means the coexistence of GHZ and W classes.}
\end{center}
\end{figure}

The Lagrange multiplier method used in the previous section to derive the region for the separable states cannot be used to derive the 
region for the bi-separable states. The reason is as follows. If we choose first qubit as a separable qubit such as 
$\ket{\psi^{bisep}} = (G_1 \otimes G_2 \otimes G_3) \ket{0} \otimes \frac{1}{\sqrt{2}} (\ket{00} + \ket{11})$, where
\begin{eqnarray}
\label{cla-1}
G_j = \left(      \begin{array}{cc}
                   A_j  &  B_j         \\
                   C_j  &  D_j         
                   \end{array}          \right),
\end{eqnarray}
the mapping from a set of the three-qubit pure states to the set of the extended GHZ-symmetric states cannot change the separability of the 
first qubit because the qubit permutation is not involved in the extended GHZ symmetry. Since, however, the definition of the bi-separable 
mixed state means a quantum state whose pure-states ensemble can be represented as only separable and bi-separable states without restriction to the 
separable qubit, the Lagrange multiplier method used in the previous section cannot be applied for deriving the region for the bi-separable
extended GHZ-symmetric states. 

In this paper, instead of the Lagrange multiplier method, we use the class-specific optimal witness operators
\begin{eqnarray}
\label{witness-1}
& & \mathcal{W}_{\mathrm{bisep\setminus sep}} = \openone
  - 4\ket{\mathrm{GHZ}_+}\!\bra{\mathrm{GHZ}_+}
  + 2\ket{\mathrm{GHZ}_-}\!\bra{\mathrm{GHZ}_-}                                           \\   \nonumber
& &\mathcal{W}_{W\setminus\mathrm{bisep}} = \frac{1}{2}\openone - \ket{\mathrm{GHZ}_+}\!\bra{\mathrm{GHZ}_+}     \\   \nonumber
& & \mathcal{W}_{\mathrm{GHZ}\setminus W}(v_0) = \frac{3}{4}\openone -
  \frac{3}{v_0^2-2v_0+4}\ \ket{\mathrm{GHZ}_+}\!\bra{\mathrm{GHZ}_+} -
  \frac{3}{v_0^2+2v_0+4}\ \ket{\mathrm{GHZ}_-}\!\bra{\mathrm{GHZ}_-},
\end{eqnarray}
which are derived in Ref.\cite{elts12-2} by using the classification of the GHZ-symmetric states. Here, we choose $v_0 = 0.981$, 
which corresponds to the fact that the optimal witness operators yield an exact classification to the Werner state
\begin{equation}
\label{werner}
\rho^W = p  \ket{\mathrm{GHZ}_+}\!\bra{\mathrm{GHZ}_+} + (1 - p) \frac{1}{8} \openone.
\end{equation}

The information the class-specific witness $\mathcal{W}_{\mathrm{A \setminus B}}$ provides is as follows. Let $\rho$ be an 
arbitrary three-qubit quantum state. If $\mbox{tr} (\mathcal{W}_{\mathrm{A \setminus B}} \rho) < 0$, this means that 
$\rho$ is in A or its higher class in the three-qubit hierarchy S $\subset$ B $\subset$ W $\subset$ GHZ. 

Using Eq. (\ref{witness-1}) and Eq. (\ref{eghz-general-form}) it is straightforward to show
\begin{eqnarray}
\label{witness-2}
& &\mbox{tr} \left[\mathcal{W}_{\mathrm{bisep\setminus sep}} \rho^{ES} \right] = \frac{3}{4} - (y_1 + y_2 + y_3) - 6 x     \\   \nonumber
& &\mbox{tr} \left[\mathcal{W}_{\mathrm{W\setminus bisep}} \rho^{ES} \right] = 
\frac{1}{2} \left[ \frac{3}{4} - (y_1 + y_2 + y_3) - 2 x \right]                                                     \\   \nonumber
& &\mbox{tr} \left[\mathcal{W}_{\mathrm{GHZ\setminus W}} \rho^{ES} \right]                                           \\    \nonumber
& &= 
\frac{3 (v_0^2 + 4)}{(v_0^2 - 2 v_0 + 4) (v_0^2 + 2 v_0 + 4)}
\left[\frac{(v_0^2 + 3) (v_0^2 + 4) - 4 v_0^2}{4 (v_0^2 + 4)} - (y_1 + y_2 + y_3) - \frac{4}{v_0^2 + 4} x \right].
\end{eqnarray}
It is worthwhile noting that Eq. (\ref{witness-2}) is not dependent on $y_j \hspace{.2cm} (j=1,2,3)$ individually, but dependent on 
$y_1 + y_2 + y_3$. 
The information we can gain from Eq. (\ref{witness-2}) is as follows. The extended GHZ-symmetric separable states should be 
confined in a polygon $(b,f,g,h)$ in Fig. 4. The extended GHZ-symmetric bi-separable states should be confined in a polygon
$(b,e,g,i)$. The extended GHZ-symmetric W states should be confined in a polygon $(a,c,d,g,j)$. Of course, all SLOCC classes 
should be distributed with obeying the three-qubit hierarchy S $\subset$ B $\subset$ W $\subset$ GHZ. This information is pictorially 
depicted in Fig. 4. The three examples discussed in the previous section can be shown to be consistent with this information, 
i.e., all green regions in Fig. 3 are contained in the polygon $(b,f,g,h)$.

There is another way, which enables us to get a rough classification of the extended GHZ-symmetric states $\rho^{ES}$. 
First, we map from $\rho^{ES}$ in Eq. (\ref{eghz-general-form}) to $\rho^S$ in Eq. (\ref{ghz-general-form}), which 
results in $\rho^S (\rho^{ES})$. Then, the parameters $x$ and $y$ of $\rho^S (\rho^{ES})$ are
\begin{eqnarray}
\label{add1}
& &x = \frac{1}{2} \left( \rho_{000,111}^{ES} + \rho_{111,000}^{ES} \right) = x                              \\   \nonumber
& &y = \frac{1}{\sqrt{3}} \left( \rho_{000,000}^{ES} + \rho_{111,111}^{ES} - \frac{1}{4} \right) = \frac{1}{\sqrt{3}} (y_1+y_2+y_3).
\end{eqnarray}
Since $\rho^S (\rho^{ES})$ should be lower class than $\rho^{ES}$, we conclude

\noindent 
(i) $\rho^{ES}$ is a GHZ class if $\rho^S (\rho^{ES})$ is a GHZ class

\noindent
(ii) $\rho^{ES}$ is a GHZ or W class if $\rho^S (\rho^{ES})$ is a W class

\noindent
(iii)  $\rho^{ES}$ is a GHZ, W, or B class if $\rho^S (\rho^{ES})$ is a B class.

\noindent
This makes a similar (but not exactly same) figure to Fig. 4.

\section{Conclusion}
In this paper we analyze the SLOCC classification of the extended GHZ-symmetric states $\rho^{ES}$, which is parametrized
by four real parameters. The condition for separable states of $\rho^{ES}$ is analytically derived (see Eq. (\ref{sep-6})). 
The higher classes such as B, W, and GHZ classes are roughly classified by making use of the class-specific optimal 
witnesses and map from extended GHZ symmetry to GHZ symmetry (see Eq. (\ref{add1})). From this analysis we guess that the 
entanglement classes of $\rho^{ES}$ are not dependent on $y_j \hspace{.2cm} (j=1,2,3)$ individually, but dependent on 
$y_1 + y_2 + y_3$ collectively. Unfortunately, we do not know how to prove our guess from the analytical ground.

The entanglement classification for the GHZ-symmetric case can be extended to the higher-qubit systems. However, analysis 
of the entanglement classes in the higher-qubit systems seems to be much more difficult than that of the three-qubit case. 
For example, the general form of the GHZ-like-symmetric states\footnote{Here, GHZ-like symmetry means a symmetry under 
(i) simultaneous flips (ii) qubit permutation (iii) qubit rotations about the $z$-axis of the form 
$$U (\phi_1, \phi_2, \phi_3) = e^{i \phi_1 \sigma_z} \otimes e^{i \phi_2 \sigma_z} \otimes e^{i \phi_3 \sigma_z}
                                \otimes e^{-i (\phi_1 + \phi_2 + \phi_3) \sigma_z}.   $$}
in four qubit system is parametrized by three real parameters in a form
\begin{eqnarray}
\label{4q-1}
& &\rho_4^S = \beta \left[ \ket{0000}\bra{1111} + \ket{1111} \bra{0000} \right]                            \\    \nonumber
& &+ \mbox{diag} \left(\alpha_1, \alpha_2, \alpha_2, \alpha_3, \alpha_2, \alpha_3, \alpha_3, \alpha_2, \alpha_2, \alpha_3, \alpha_3, \alpha_2, 
                    \alpha_3, \alpha_2, \alpha_2, \alpha_1   \right)
\end{eqnarray}
with $\alpha_1 + 4 \alpha_2 + 3 \alpha_3 = \frac{1}{2}$. Therefore, total set of the GHZ-like symmetric states should be 
represented by three-dimensional volume in the parameter space. Furthermore, although the entanglement classification of the 
four-qubit pure system is treated in several papers\cite{fourP,fourP-2,fourP-3,fourP-4,fourP-5,fourP-6}, their results can be confusing and 
seemingly contradictory. The worst thing is that the entanglement classes of the four-qubit mixed system are not well understood 
so far and it is not clear whether or not they obey the linear hierarchy. We hope to revisit this issue in the future.

{\bf Acknowledgement}:
This research was supported by the Kyungnam University Research Fund, 2013.


\begin{thebibliography}{99}
\bibitem{horodecki09} R. Horodecki, P. Horodecki, M. Horodecki, and K. Horodecki, {\it Quantum Entanglement}, Rev. Mod. Phys. 
{\bf 81} (2009) 865 [quant-ph/0702225] and references therein.
\bibitem{text} M. A. Nielsen and I. L. Chuang, Quantum Computation and Quantum Information (Cambridge
University Press, Cambridge, England, 2000).
\bibitem{teleportation} C. H. Bennett, G. Brassard, C. Cr´epeau, R. Jozsa, A. Peres and W. K. Wootters, {\it Teleporting
an Unknown Quantum State via Dual Classical and Einstein-Podolsky-Rosen Channles}, Phys.Rev. Lett. {\bf 70} (1993) 1895.
\bibitem{superdense} C. H. Bennett and S. J. Wiesner, {\it Communication via one- and two-particle operators on
Einstein-Podolsky-Rosen states}, Phys. Rev. Lett. {\bf 69} (1992) 2881.
\bibitem{clon} V. Scarani, S. Lblisdir, N. Gisin and A. Acin, {\it Quantum cloning}, Rev. Mod. Phys. {\bf 77} (2005)
1225 [quant-ph/0511088] and references therein.
\bibitem{cryptography} A. K. Ekert, {\it Quantum Cryptography Based on Bell’s Theorem}, Phys. Rev. Lett. {\bf 67} (1991)
661.
\bibitem{computer} G. Vidal, {\it Efficient classical simulation of slightly entangled quantum computations}, Phys. Rev.
Lett. {\bf 91} (2003) 147902 [quant-ph/0301063].
\bibitem{bennet00} C. H. Bennett, S. Popescu, D. Rohrlich, J. A. Smolin, and A. V. Thapliyal, {\it Exact and asymptotic measures
of multipartite pure-state entanglement}, Phys. Rev. {\bf A 63} (2000) 012307.
\bibitem{dur00} W. D\"{ur}, G. Vidal and J. I. Cirac, {\it Three qubits can be entangled in two inequivalent ways},
Phys.Rev. A {\bf 62} (2000) 062314.
\bibitem{threeM} A. Ac\'{i}n, D. Bru\ss, M. Lewenstein, and A. Sanpera, {Classification of Mixed Three-Qubit States}, 
Phys. Rev. Lett. {\bf 87} (2001) 040401.
\bibitem{elts12-2} C. Eltschka and J. Siewert, {\it Optimal witnesses for three-qubit entanglement from Greenberger-Horne-Zeilinger symmetry},
arXiv:1204.5451 (quant-ph).
\bibitem{fourP} F. Verstraete, J. Dehaene, B. De Moor, and H. Verschelde, {\it Four qubits can be entangled in nine different ways}, 
Phys. Rev. {\bf A 65} (2002) 052112.
\bibitem{concurrence1} W. K. Wootters, {\it Entanglement of Formation of an Arbitrary State of Two Qubits}, Phys. Rev.
Lett. {\bf 80} (1998) 2245 [quant-ph/9709029]. 
\bibitem{ckw} V. Coffman, J. Kundu and W. K. Wootters, {\it Distributed entanglement}, Phys. Rev. {\bf A 61} (2000) 052306 [quant-ph/9907047].
\bibitem{elts12-1} C. Eltschka and J. Siewert, {\it Entanglement of Three-Qubit Greenberger-Horne-Zeilinger-Symmetric States}, 
Phys. Rev. Lett. {\bf 108} (2012) 020502.
\bibitem{siewert12-1} J. Siewert and C. Eltschka, {\it Quantifying Tripartite Entanglement of Three-Qubit Generalized Werner States}, 
Phys. Rev. Lett. {\bf 108} (2012) 230502.
\bibitem{fourP-2} L. Lamata, J. Le\'on, D. Salgado, and E. Solano, {\it Inductive entanglement of four qubits under stochastic local 
operations and classical communication}, Phys. Rev. {\bf A 75} (2007) 022318.
\bibitem{fourP-3} Y. Cao and A. M. Wang, {\it Discussion of the entanglement classification of a 4-qubit pure state}, 
Eur. Phys. J. {\bf D 44} (2007) 159.
\bibitem{fourP-4} D. Li, X. Li, H. Huang, and X. Li, Quantum Inf. Comput. {\bf 9} (2009) 0778.
\bibitem{fourP-5} S. J. Akhtarshenas and M. G. Ghahi, {\it Entangled graphs: A classification of four-qubit entanglement}, 
arXiv:1003.2762 (quant-ph).
\bibitem{fourP-6} L. Borsten, D. Dahanayake, M. J. Duff, A. Marrani, and W. Rubens, {\it Four-Qubit Entanglement Classification from 
String Theory}, Phys. Rev. Lett. {\bf 105} (2010) 100507.















\end{thebibliography}
\end{document}